\documentclass[reprint
,amsmath
,amssymb
,aps
,pre
,superscriptaddress
]{revtex4-2}

\usepackage[utf8]{inputenc}
\usepackage[T1]{fontenc}
\usepackage[pdftex]{graphicx}
\usepackage{xurl}
\usepackage{orcidlink}
\usepackage{nicefrac}
\usepackage{setspace}
\usepackage{placeins}
\usepackage{todonotes}
\usepackage{algorithm}
\usepackage[noend]{algpseudocode}

\usepackage{siunitx}
\DeclareSIUnit{\EUR}{€}
\DeclareSIUnit{\billion}{\text{bn.}}

\begin{document}

\title{
Cascading Failures and Critical Infrastructures in Future Renewable European Power Systems
}

\author{Maurizio Titz \orcidlink{0000-0002-0249-6244}}
\email{m.titz@fz-juelich.de}
\affiliation{Forschungszentrum J\"ulich, Institute of Climate and Energy Systems (ICE-1), 52428 J\"ulich, Germany}
\affiliation{Institute for Theoretical Physics, University of Cologne, K\"oln, 50937, Germany}

\author{Franz Kaiser \orcidlink{0000-0002-7089-2249}}
\affiliation{Institute for Theoretical Physics, University of Cologne, K\"oln, 50937, Germany}
    
\author{Johannes Kruse \orcidlink{0000-0002-3478-3379}}
\affiliation{Institute for Theoretical Physics, University of Cologne, K\"oln, 50937, Germany}

\author{Philipp C. B\"ottcher\orcidlink{0000-0002-3240-0442}}
  \affiliation{Forschungszentrum J\"ulich, Institute of Climate and Energy Systems (ICE-1), 52428 J\"ulich, Germany}

\author{Jan Lange \orcidlink{0009-0009-5985-4848}}
\affiliation{Forschungszentrum J\"ulich, Institute of Climate and Energy Systems (ICE-1), 52428 J\"ulich, Germany}
\affiliation{Institute for Theoretical Physics, University of Cologne, K\"oln, 50937, Germany}
    
\author{Martha Frysztacki\orcidlink{0000-0002-0788-1328}}
\affiliation{Institute for Automation and Applied Informatics, Karlsruhe Institute of Technology, 76344 Eggenstein-Leopoldshafen, Germany}
\affiliation{Open Energy Transition, 95448 Bayreuth, Germany}

\author{Dominic Hewes}
\affiliation{TenneT TSO GmbH, Bernecker Str. 70, 95448 Bayreuth, Germany}

\author{Michael Orlishausen}
\affiliation{TenneT TSO GmbH, Bernecker Str. 70, 95448 Bayreuth, Germany}

\author{Mark Thiele}
\affiliation{TenneT TSO GmbH, Bernecker Str. 70, 95448 Bayreuth, Germany}

\author{Tom Brown\orcidlink{0000-0001-5898-1911}}
\affiliation{Institute for Automation and Applied Informatics, Karlsruhe Institute of Technology,
76344 Eggenstein-Leopoldshafen, Germany} 
\affiliation{Department of Digital Transformation in Energy Systems, Technische Universität Berlin, Einsteinufer 25 (TA 8), 10587 Berlin, Germany}

\author{Dirk Witthaut\orcidlink{0000-0002-3623-5341}}%
\email{d.witthaut@fz-juelich.de}
\affiliation{Forschungszentrum J\"ulich, Institute of Climate and Energy Systems (ICE-1), 52428 J\"ulich, Germany}
\affiliation{Institute for Theoretical Physics, University of Cologne, K\"oln, 50937, Germany}
\date{\today}

\begin{abstract}
    The world's power systems are undergoing a rapid transformation, shifting away from carbon-intensive power generation to renewable sources. 
    As a result, electricity is being transported over ever longer distances, while the intrinsic system inertia provided by thermal power plants decreases. 
    Together, these developments raise the probability of cascading line failures and reduce the stability of the system after a system split.
    In this article, we assess the risk of cascading failures and system splits in the European power grid for different carbon reduction scenarios. 
    We analyze the most likely and most dangerous splits, and identify critical transmission infrastructures and we discuss potential countermeasures that can address the problem of cascades. 
    Our results show that while the risks of splits causing power failures rises with decarbonization, it can be mitigated cost efficiently.
\end{abstract}

\maketitle

\section{Introduction}

Mitigating climate change requires a comprehensive transformation of the energy system towards renewable electricity~\cite{rogelj2015energy}. This shift has major implications for system operation: wind and solar generation is variable~\cite{Staffell2018}, and suitable sites are often distant from demand centres, increasing transmission requirements~\cite{rodriguez2014transmission,hallerDecarbonizationScenariosEU2012,titz2024identifying}. Given the critical importance of secure electricity supply, it is essential to understand how this transformation affects power system stability and which measures are required to maintain reliability~\cite{xu2024resilience}.

Most power supply disturbances originate in distribution grids and have limited local consequences~\cite{campbell2012weather}. Large-scale blackouts are rare but can have severe societal impacts~\cite{tab2011}. They are typically triggered by the failure of one or more transmission elements, followed by cascading outages~\cite{pourbeik2006anatomy,sturmer2024increasing}. 
In extreme cases, the system splits into electrically isolated areas prone to instability, as observed in the Italian and Northeastern blackouts in 2003~\cite{andersson2005causes}, the Western European blackout in 2006~\cite{UCTE07}, and further splits in 2021~\cite{entsoe2021a,entsoe2021b}. A system split also occurred during the 2025 Iberian blackout, although it was triggered primarily by voltage stability issues rather than cascading line overloads~\cite{entsoe2025}. These events raise a key question: how does the risk of cascading failures evolve during the energy transition, and how can it be mitigated?

Many post-contingency stabilisation mechanisms rely on system inertia~\cite{machowski_power_2008,tielens_relevance_2016}. Replacing inertia-providing conventional generation with renewables may therefore challenge stability~\cite{milano_foundations_2018}. At the same time, increasing power transmission increases the likelihood of overloads and cascading failures~\cite{kaiser_universal_2021}. The combination of the two can be particularly critical. 
Market coupling and spatially concentrated renewable generation increase grid utilisation and inter-area transfers, thereby amplifying the risk of cascading failures.
Following a split, these transfers translate into substantial power imbalances that are harder to stabilise under low inertia. In such cases, operators rely on measures such as load shedding~\cite{de_boeck_review_2016,entsoe_P5}, which are not always sufficient to prevent a large-scale blackout.
Importantly, these risks are largely determined by how renewable generation is integrated into the grid. Empirical analyses suggest that high shares of wind and solar generation do not necessarily increase blackout vulnerability~\cite{zhao2024impacts}.

\begin{figure}[tb]
    \centering
    \includegraphics[width=\columnwidth]{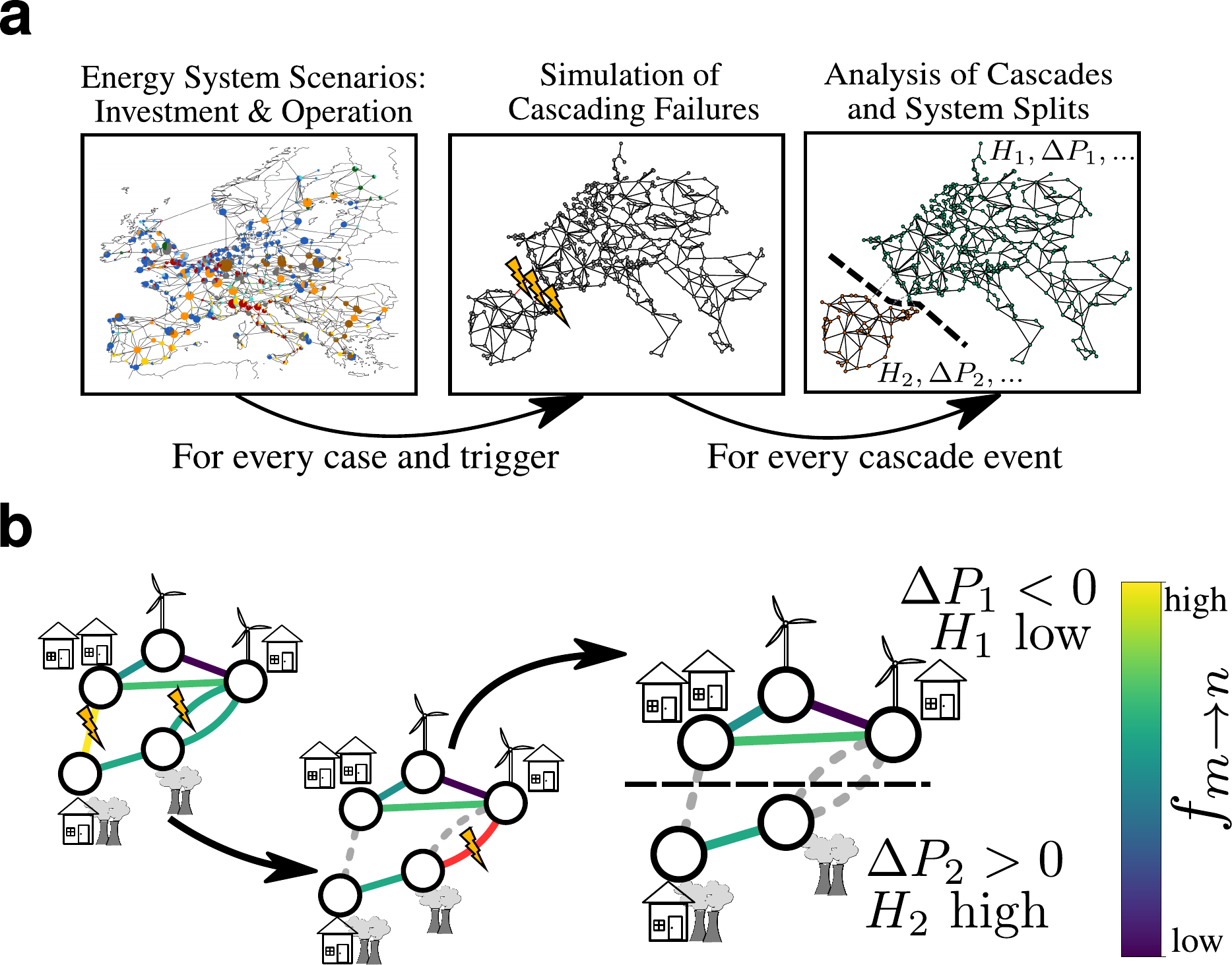}
    \caption{
    \textbf{Schematic of the analysis.}
    \textbf{a,} The power system model PyPSA provides scenario data for different CO$_2$ targets and an $(N-1)$-secure dispatch at each time step. For every scenario and time step, we simulate all double transmission failures. These may trigger cascades and eventual system splits, which are analysed statistically.
    \textbf{b,} Schematic of a cascading failure in a power grid. An initial outage of two transmission lines (lightning bolts) redistributes real power flows $f_{m\rightarrow n}$ and may overload additional lines. The system can split into components with generation–load imbalance $\Delta P_c$. Splits are most critical when $\Delta P_c$ is large and the effective inertia $H_c$ is low.
    }
    \label{fig:scheme}
\end{figure}

Here we quantify how the energy transition alters the likelihood and impacts of large-scale blackouts in Europe. Using cost-optimal decarbonisation scenarios, we simulate cascading failures and system splits across a wide range of operating conditions (Fig.~\ref{fig:scheme}). This enables a statistical assessment of key risk factors, the probability and consequences of system splits, critical infrastructures, and potential countermeasures.

\section{Power System Scenarios and Simulations}

Our analysis is based on scenarios generated with the open energy system model PyPSA-Eur~\cite{horsch2018pypsa}, which co-optimises generation, storage and dispatch under transmission constraints. For a given CO$_2$ target, total annualised system costs are minimised subject to demand satisfaction and an emissions cap relative to 1990 levels. The model yields installed capacities and dispatch in 3-hour resolution for all generation and storage technologies. Transmission capacities are fixed, and power flows are represented using the linear (DC) approximation~\cite{Wood14}. Dispatch is optimized under the constraint that $N-1$ security is satisfied in every time step, ensuring that no outage of a single transmission line leads to secondary overloads~\cite{Wood14}. The explicit enforcement of operational security constraints is uncommon in large-scale energy system optimisation models and ensures that cascading failures are evaluated from system states that comply with standard $N-1$ security.

\begin{figure*}[t]
    \begin{center}
    \includegraphics[width=1.0\textwidth]{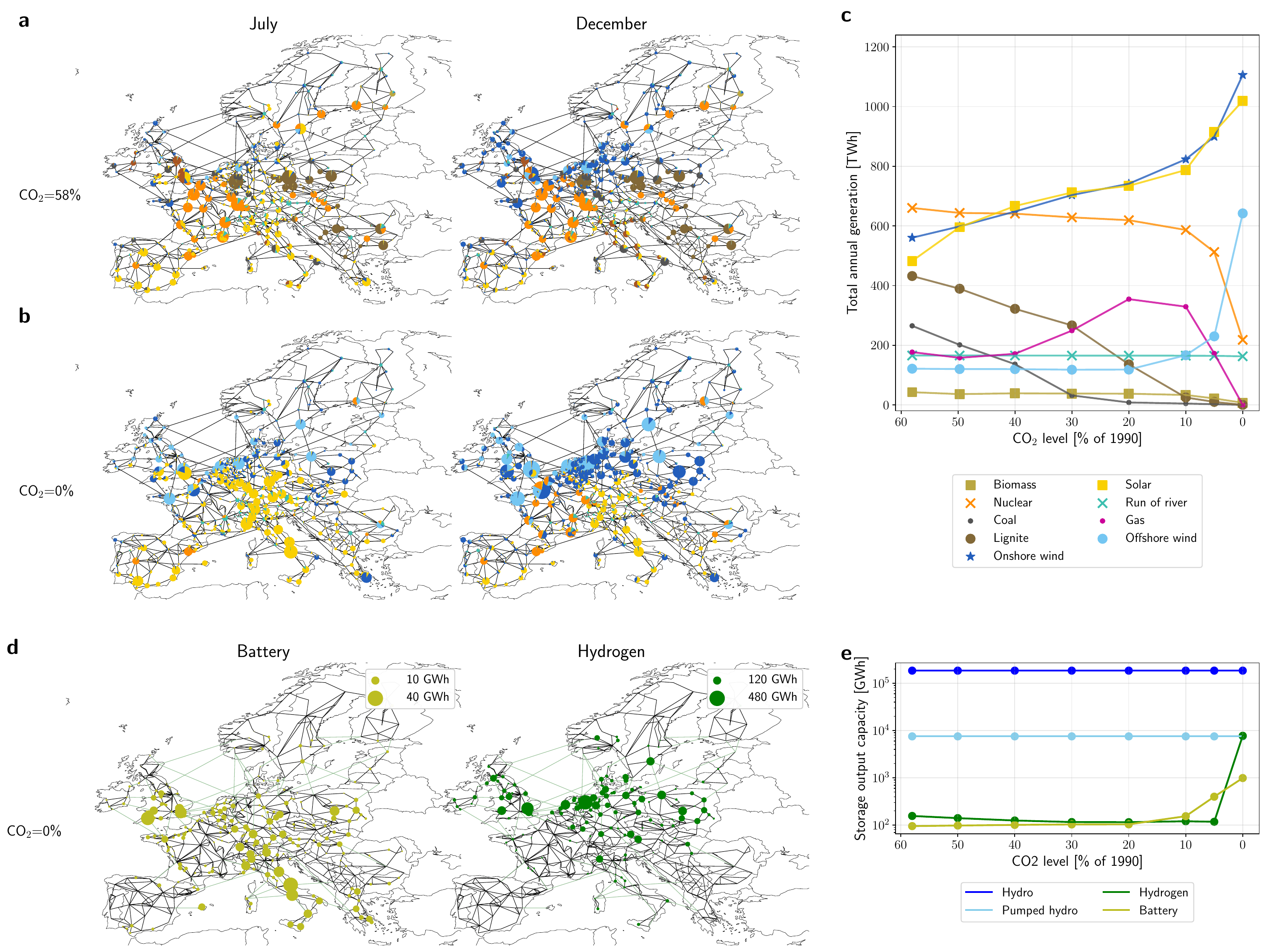}
    \end{center}
    \caption{
    \textbf{Scenarios for the decarbonisation of the European power system.}
    \textbf{a,b,} Spatial distribution of cumulative generation in July and December at CO$_2$ levels of $58\%$ and $0\%$ relative to 1990. Disk size indicates monthly electricity generation per node. The $58\%$ scenario approximates the 2022 European system and serves as a reference. At $0\%$, generation is dominated by wind, solar, nuclear and hydropower, with pronounced seasonal and geographic variation.
    \textbf{c,} Annual generation by technology as a function of CO$_2$ level. Fossil generation is progressively replaced by wind and solar.
    \textbf{d,} Spatial distribution of battery and H$_2$ storage output capacity in the $0\%$ scenario. Batteries cluster near solar generation, while H$_2$ storage concentrates near wind-dominated regions.
    \textbf{e,} Total installed storage output capacity (logarithmic scale). H$_2$ storage increases sharply at $0\%$ as gas plants are no longer available to balance prolonged wind deficits, while battery capacity grows strongly beyond 20\% decarbonisation.
    }
    \label{fig:generation_mix_storage}
\end{figure*}

The resulting scenarios are summarised in Fig.~\ref{fig:generation_mix_storage}. The reference case (58\% CO$_2$) is chosen because it closely reflects current emission levels and resembles today’s system. In the fully decarbonised scenario, generation is dominated by wind and solar power, characterised by strong temporal variability and spatial heterogeneity. Wind generation prevails in northern Europe, especially around the North Sea, while solar capacities concentrate in southern regions. Their spatial distribution is further shaped by transmission constraints and load distribution. For instance, this leads to large installations of PV along an axis ranging from Benelux over Germany to Northern Italy 
east of a grid bottleneck. 
At intermediate CO$_2$ levels, gas plants provide flexibility. At 0\% emissions, hydrogen storage becomes essential to bridge prolonged periods of low renewable output, and its capacity increases by two orders of magnitude compared to higher-emission scenarios. Batteries provide intra-day flexibility for solar generation, with capacities rising markedly once CO$_2$ emissions fall below 20\%.
Nuclear utilisation declines gradually at first, primarily during hours of high solar output.
Below 20\% emissions, it drops markedly as battery storage enables solar generation to be shifted across the day.

Cascading failures and system splits are simulated as illustrated in Fig.~\ref{fig:scheme}. For each scenario and time step, we evaluate all $N-2$ contingencies by removing pairs of transmission circuits. Power flows are recalculated and overloaded lines are iteratively removed until no further violations occur or the system separates into disconnected components. In the latter case, we analyse the resulting system state and potential blackouts.

Overall, we simulate around $1.1\times10^9$ $N-2$ contingencies per CO$_2$ level. All results are conditioned on the occurrence of such double contingencies. Although simultaneous line failures are rare, this systematic sampling enables a statistical assessment of structural vulnerability. To render this large-scale analysis tractable, we adopt a coarse-grained network representation and a steady-state power flow approximation (see Appendix). Rather than reproducing detailed transient dynamics of individual events, we quantify how structural changes in generation, storage and transmission alter susceptibility to cascading failures; detailed dynamic case studies are available elsewhere~\cite{ippolito_analysis_2021,kannan_frequency_2020}.

\section{Risk factors for cascading failures and Blackouts}

\begin{figure}
    \centering
    \includegraphics[width=\linewidth]{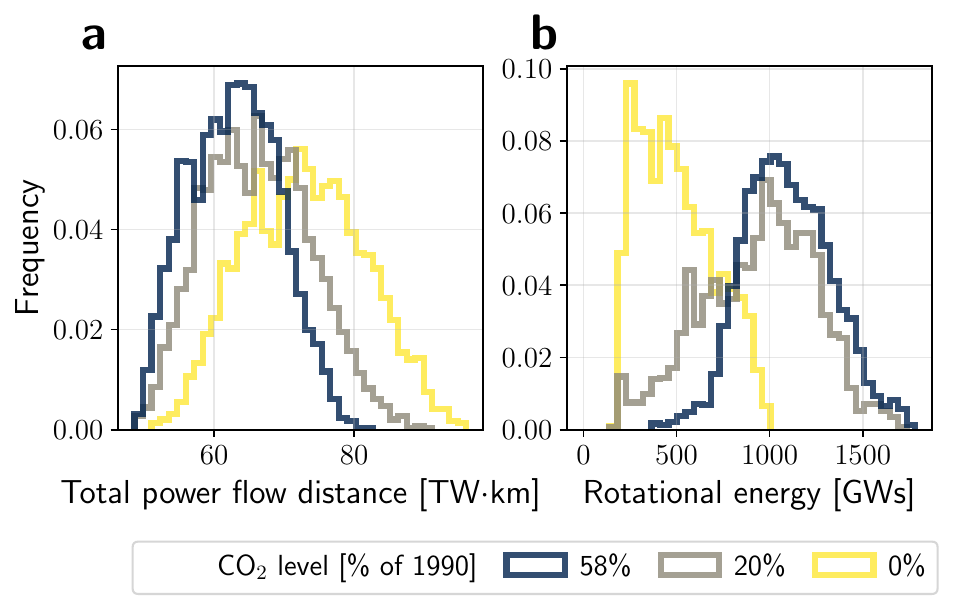}
    \caption{
    \textbf{Pre-outage risk factors.}
    \textbf{a,} Total power flow distance across emission levels. Higher values indicate longer average transfer distances and increased grid loading. The increase is constrained by fixed transmission capacities, which are not optimised.
    \textbf{b,} Total rotational energy of the power system. As decarbonisation progresses, inertia declines sharply, with very low values prevailing in the fully decarbonised scenario.
    }
    \label{fig:flow_inertia}
\end{figure}

Favourable locations for wind and solar generation are often distant from major load centres, increasing transmission requirements. To quantify overall grid loads, we define the total power flow distance as the product of active power flow and line length, summed over all transmission lines (Fig.~\ref{fig:flow_inertia}\textbf{a}). As emissions decline, the distribution shifts towards higher values; in the zero-emission scenario, the maximum increases by 17\% relative to the reference case. Variability also grows, reflecting stronger seasonal and synoptic fluctuations in renewable generation. Larger flow distances indicate heavier network utilisation and therefore a higher propensity for cascading failures.

Whether a cascade results in a blackout depends on the grid response. A power imbalance causes the grid frequency to deviate at a rate inversely proportional to system inertia. If the rate of change of frequency is too high, primary control and load shedding may fail to stabilise the system. In AC systems, inertia is provided by the kinetic energy of synchronous machines. We quantify available inertia at time $t$ via the rotational energy
\begin{equation}
E_{\rm kin}(t) = \sum_i S_{B,i} H_i \Theta_i(t),
\end{equation}
where $S_{B,i}$ is the apparent power rating and $H_i$ the inertia constant of plant $i$. Conventional thermal and nuclear units typically exhibit inertia constants between 3 and 6 seconds~\cite[p.~24]{entsoe2020ROCOF}. The factor $\Theta_i(t)$ indicates whether plant $i$ is online. As full unit commitment is infeasible at continental scale, we approximate this by assuming a unit is online when its output exceeds 5\% of rated capacity (cf.~Appendix). Load contributes weakly to the inertia with $H=0.4$ seconds~\cite{Stenzel22}, in particular through directly connected rotating machines.

Decarbonisation drastically reduces available inertia (Fig.~\ref{fig:flow_inertia}\textbf{b}).
While kinetic energy never falls below $367~\text{GWs}$ in the reference scenario (58\% CO$_2$), it can decline to $180~\text{GWs}$ at $0\%$. In this case, inertia is mainly provided by nuclear and hydro power plants and load. This structural reduction in inertia increases the vulnerability of split subsystems to instability and blackout.

\section{The risk of cascading failures and system splits}

\begin{figure*}[tb!]
    \begin{center}
    \includegraphics[width=.9\linewidth]{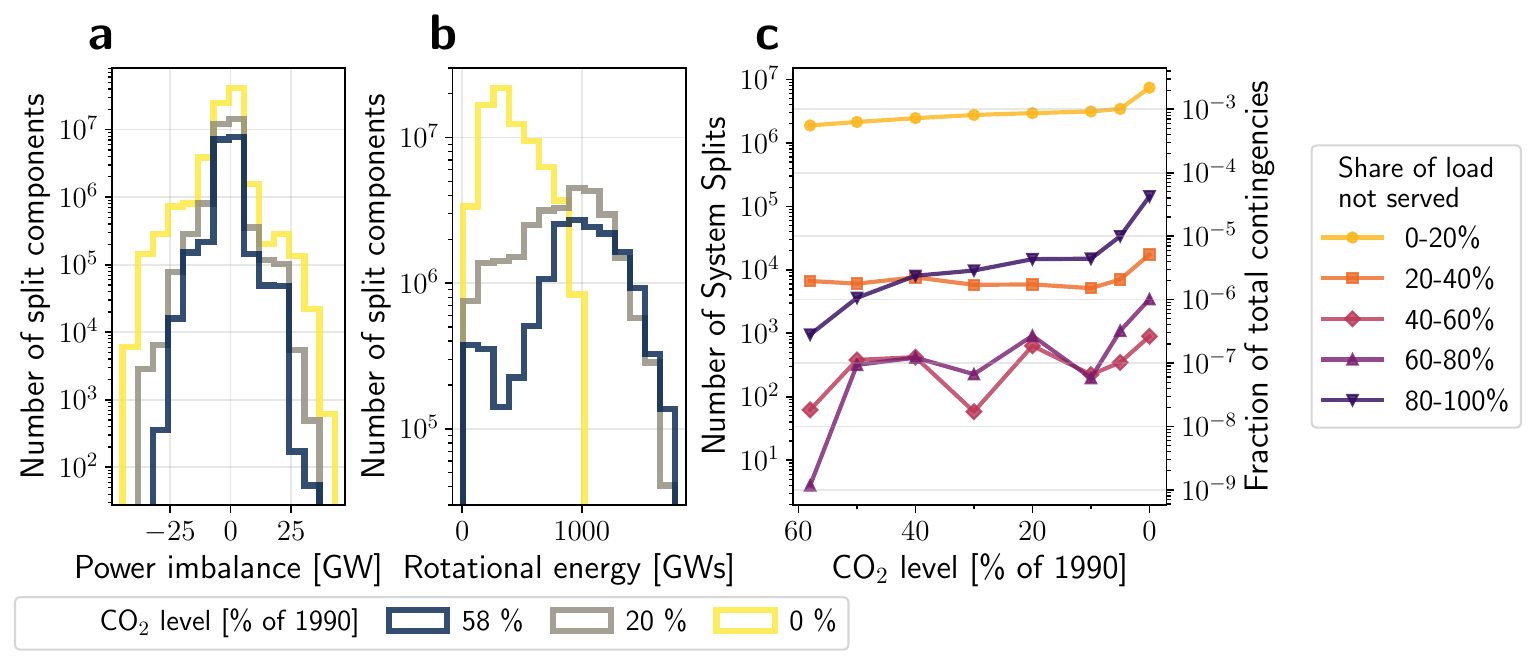}
    \end{center}
    \caption{
    \textbf{Statistical assessment of system splits triggered by cascading failures.}
    \textbf{a,} Distribution of power imbalance $\Delta P_c$ across split components. The overall shape remains similar across emission levels, but larger imbalances, particularly negative ones, become more frequent.
    \textbf{b,} Distribution of rotational energy $E_{{\rm kin},c}$ across split components. Very low-inertia splits become substantially more likely in the fully decarbonised system. Components carrying less than 10\% of total load are excluded for clarity, as they exhibit low inertia by construction.
    \textbf{c,} Number of split events as a function of CO$_2$ level, grouped by share of load not served. The secondary axis indicates the probability conditional on an $N-2$ failure. While all outage sizes increase as emissions decline, the rise is most pronounced for the largest events corresponding to global severe splits.
    }
    \label{fig:split_statistics} 
\end{figure*}

Cascading failures can split the grid into disconnected components with large power imbalances and low inertia. The severity of such splits depends on the resulting imbalance and the available inertia within each component. As CO$_2$ emissions decrease, split components exhibit moderately larger power imbalances but substantially lower inertia (Fig.~\ref{fig:split_statistics}). Most components exhibit $|\Delta P_c| \le 10\,\si{GW}$, while at 0\% emissions rare events reach up to $20\,\si{GW}$. In contrast, the number of components with inertia below $500\,\si{GWs}$ increases by more than a factor of 30 as CO$_2$ approaches zero.

The severity of a split is quantified by the \emph{Rate of Change of Frequency (RoCoF)}~\cite{broderick_rate_nodate}. For a disconnected component $c$ with imbalance $\Delta P_c$ and rotational energy $E_{{\rm kin},c}$, the signed RoCoF is~\cite{ulbigImpactLowRotational2014}
\begin{align}
    \text{RoCoF}_c = \frac{f_0 \Delta P_c}{2 E_{{\rm kin},c}},
    \label{eq:rocof}
\end{align}
where $f_0 = 50\,\si{Hz}$. If RoCoF is too large, emergency measures such as load shedding~\cite{de_boeck_review_2016,entsoe_P5} may be insufficient to stabilise a component. To quantify the overall risk for blackouts, we assume that a component experiences a complete blackout if $|\text{RoCoF}(c)| > 1 \ \text{Hz/s}$~\cite{entsoe2020ROCOF}.

Without mitigation, the number of partial and complete blackouts increases by roughly one order of magnitude when emissions approach zero. For further analysis, we group events by the fraction of load not served (Fig.~\ref{fig:split_statistics}\textbf{c}). Small events (loss \textless20\%) dominate in all scenarios. Events with load not served above 80\% typically correspond to near-complete blackouts and align with the definition of ``global severe splits'' (GSS) in~\cite{entso-e_project_2025}. While still being rare, their frequency of occurrence rises by around two orders of magnitude at 0\% CO$_2$ compared to the 58\% reference scenario.

\section{Likely and severe system splits}

\begin{figure*}[tb]
    \begin{center}
    \includegraphics[width=\textwidth]{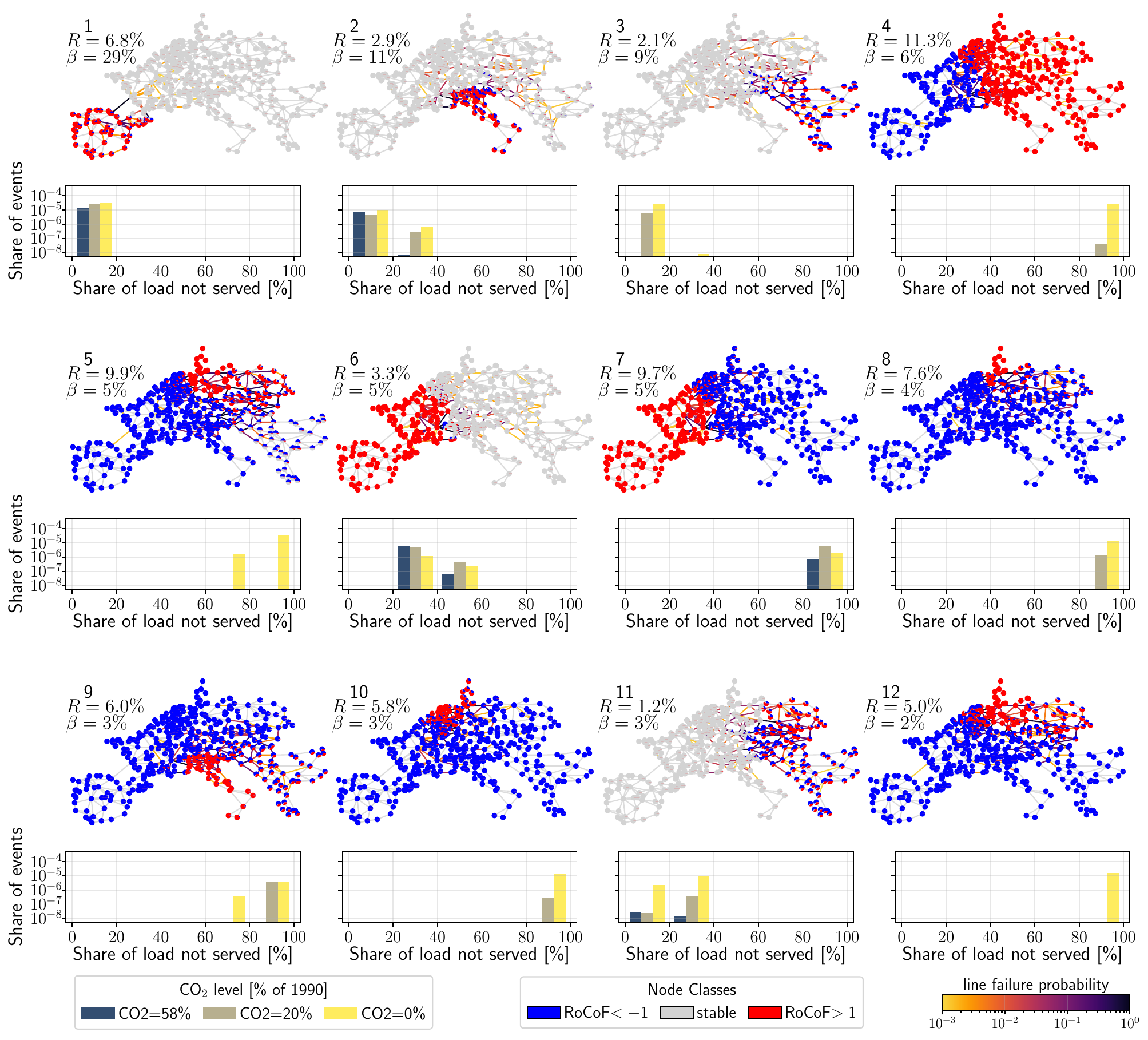}
    \end{center}
    \caption{
    \textbf{Characteristic geographic patterns of system splits.}
    Split events with more than 10\% load not served are clustered by blackout pattern (see Appendix). For each cluster, $R$ denotes its contribution to total load not served and $\beta$ its share of all considered blackouts. The 12 most frequent clusters (highest $\beta$) are shown. 
    For each cluster, the maps displays the probability that a node experiences a blackout due to positive or negative RoCoF violation or remains stable (pie charts), and the probability of transmission corridor failure (colour scale). Cut-set edges are omitted, as they connect nodes with different outcomes by construction.
    Histograms below each map show the distribution of total lost-load share within the cluster, normalised by the number of trigger events, for each CO$_2$ level.
    }
    \label{fig:cluster} 
\end{figure*}

The overall risk of system splits increases under decarbonisation --- but which patterns dominate and which are most dangerous? To answer this, we cluster all split events by geographic similarity and identify twelve main clusters covering approximately 82\% of all events (Fig.~\ref{fig:cluster}). Most frequent splits align with well-defined transmission bottlenecks, while deep decarbonisation introduces additional and more diffuse patterns.

Several clusters coincide with historically known vulnerabilities. Clusters~2 and 9 capture splits at Italy’s Alpine borders; a similar event triggered the Italian blackout in 2003~\cite{andersson2005causes}. In low-emission scenarios, such splits can lead to Europe-wide blackouts due to large negative RoCoF violations when Italian solar generation cannot be exported. Cluster~1 reflects the isolation of the Iberian Peninsula, as observed in 2021 and 2025~\cite{entsoe2021b,entsoe2025}, typically producing positive RoCoF violations from excess solar generation. Clusters~3 and partially cluster~9 and 11 correspond to Balkan-region separations similar to events in 2021~\cite{entsoe2021a}, with both positive and negative RoCoF outcomes. Cluster~5 resembles the 2006 system split through Germany and southeastern Europe~\cite{UCTE07}.

Decarbonisation gives rise to new risk patterns. Clusters~8 and 10 isolate regions around the North Sea and Denmark, where high wind generation can produce positive RoCoF violations. Although this bottleneck is well documented~\cite{titz2024identifying}, it has not yet resulted in a major split. Clusters~5 and 11 highlights vulnerabilities around Poland, where lignite generation is replaced by wind in low-emission scenarios. The increased seasonal imports and exports expose the weak connectivity along its borders. Clusters~4, 6 and 7 represent East–West separations along the French–German axis, indicating structurally central bottlenecks even without strong geographical barriers.

The impact of a split depends critically on the pre-contingency import–export balance of the separated region and the available inertia.
If a small exporting region becomes isolated, the resulting blackout is typically confined locally (e.g.~clusters~1–3). 
By contrast, in low-emission scenarios, the sudden loss of imports can destabilise the remaining system and trigger a large-scale blackout (e.g.~cluster 9).
The likelihood of such system-wide events increases strongly at very low CO$_2$ levels.

Overall, split locations are primarily shaped by transmission bottlenecks and the community structure of the grid, whereas blackout severity is determined by generation distribution and inertia. With rising shares of renewables and unchanged transmission infrastructure, splits become more extensive and severe, substantially increasing the risk of large-scale blackouts. 

\section{Identifying critical Infrastructures}

\begin{figure*}
    \centering
    \includegraphics[width=\linewidth]{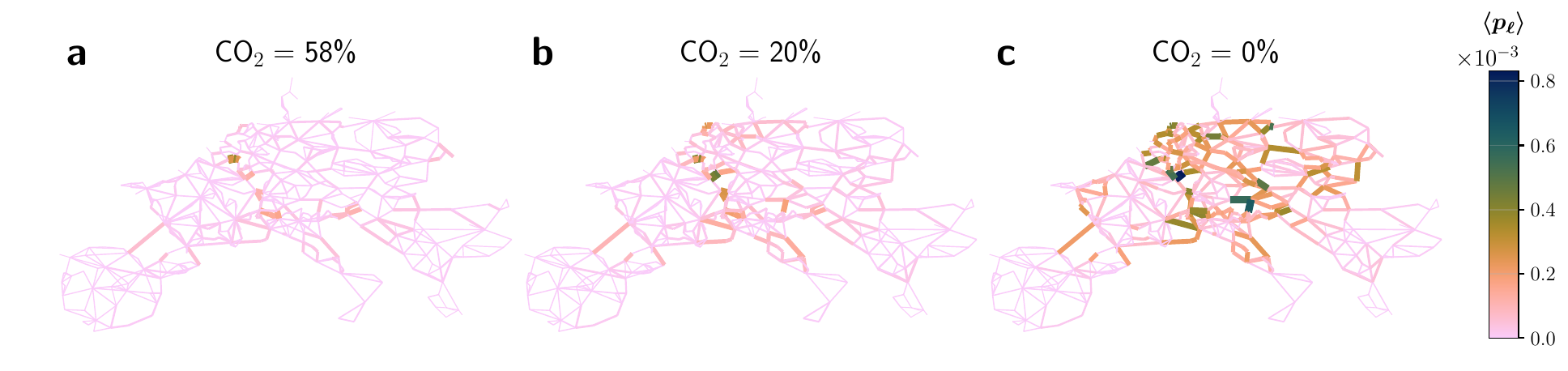}
    \caption{    
    \textbf{Conditional probability of transmission lines participating in cascading failures,} either as primary triggers or as secondary outages, across different CO${}_2$ levels. Results are conditioned on the occurrence of a $N-2$ transmission line failure.
    \textbf{a,} At the reference level (58\% of 1990 emissions), four critical regions dominate: the Germany–France interface, the Spain–France interconnection, the Balkan corridor, and Italy’s connections to the rest of the grid.
    \textbf{b,} At 20\% emissions, these regions intensify and additional vulnerabilities emerge near the North and Baltic Seas.
    \textbf{c,} At 0\% emissions, highly critical lines expand markedly, particularly across Central Europe.
    }
    \label{fig:line_failures}
\end{figure*}

To identify critical transmission infrastructures, we evaluate for each line $\ell$ the probability $\langle p_\ell \rangle$ of being involved in a cascade, either as a triggering or secondary failure, conditioned on the occurrence of an $N-2$ contingency (Fig.~\ref{fig:line_failures}), following a similar approach to Ref.~\cite{yang2017small}. Two main trends emerge: existing bottlenecks remain critical and expand spatially, and new high-risk corridors appear when emissions approach zero.

In the reference scenario, the most vulnerable lines are concentrated in Belgium and along the France-Germany border. Additional critical regions include the Alpine corridor, the Iberian and Balkan bottlenecks, and eastern Poland.

At 20\% CO$_2$, critical lines cluster predominantly in Germany, Belgium, the Netherlands and along the France–Germany axis, reflecting intensified north–south transfers from North Sea wind generation as well as substantial east–west transit flows across central Europe. The Alpine region remains a secondary hotspot.

At 0\% CO$_2$, vulnerabilities intensify and spread. All previously critical regions become more pronounced, and a new high-risk corridor emerges from northwestern Germany through eastern Germany and Austria towards northern Italy and Slovenia. This corridor closely aligns with the path of the 2006 European blackout~\cite{UCTE07}. Notably, many critical lines are located within densely meshed areas rather than at obvious geographic bottlenecks.

\section{Mitigating the risk of severe system splits}

We evaluate two countermeasures targeting the main risk drivers: declining inertia and increasing grid loading.

First, we assess the deployment of additional inertia, e.g.~via virtual inertia~\cite{fang2017distributed} or synchronous condensers~\cite{nguyen2018combination}, to limit RoCoF following a split and thereby prevent global severe splits where at least 80\% of load is lost (Fig.~\ref{fig:mitigation}a). Using a heuristic optimisation scheme (see Appendix), we determine cost-effective placement strategies and quantify the resulting reduction in the expected number of GSS across all $N-2$ contingencies. Annualised costs are estimated using prices from the German market-based procurement scheme for inertia (2026-2028)~\cite{noauthor_marktgestutzte_nodate}. At 20\% CO$_2$, approximately $318 \, \si{GWs}$ of additional inertia are required to reduce the expected number of GSS to the reference level. The additional inertia concentrates in the central grid, reflecting the objective of preventing large-scale rather than local blackouts (Fig.~\ref{fig:mitigation}). In the fully decarbonised system, more than $1000 \, \si{GWs}$ are needed, a value comparable to the median of the rotational inertia in the reference scenario. The associated annualised cost of roughly $1.1$ billion Euro is modest compared to planned transmission investments of several tens of billions per year in Germany.

\begin{figure*}
\includegraphics[width=\linewidth]{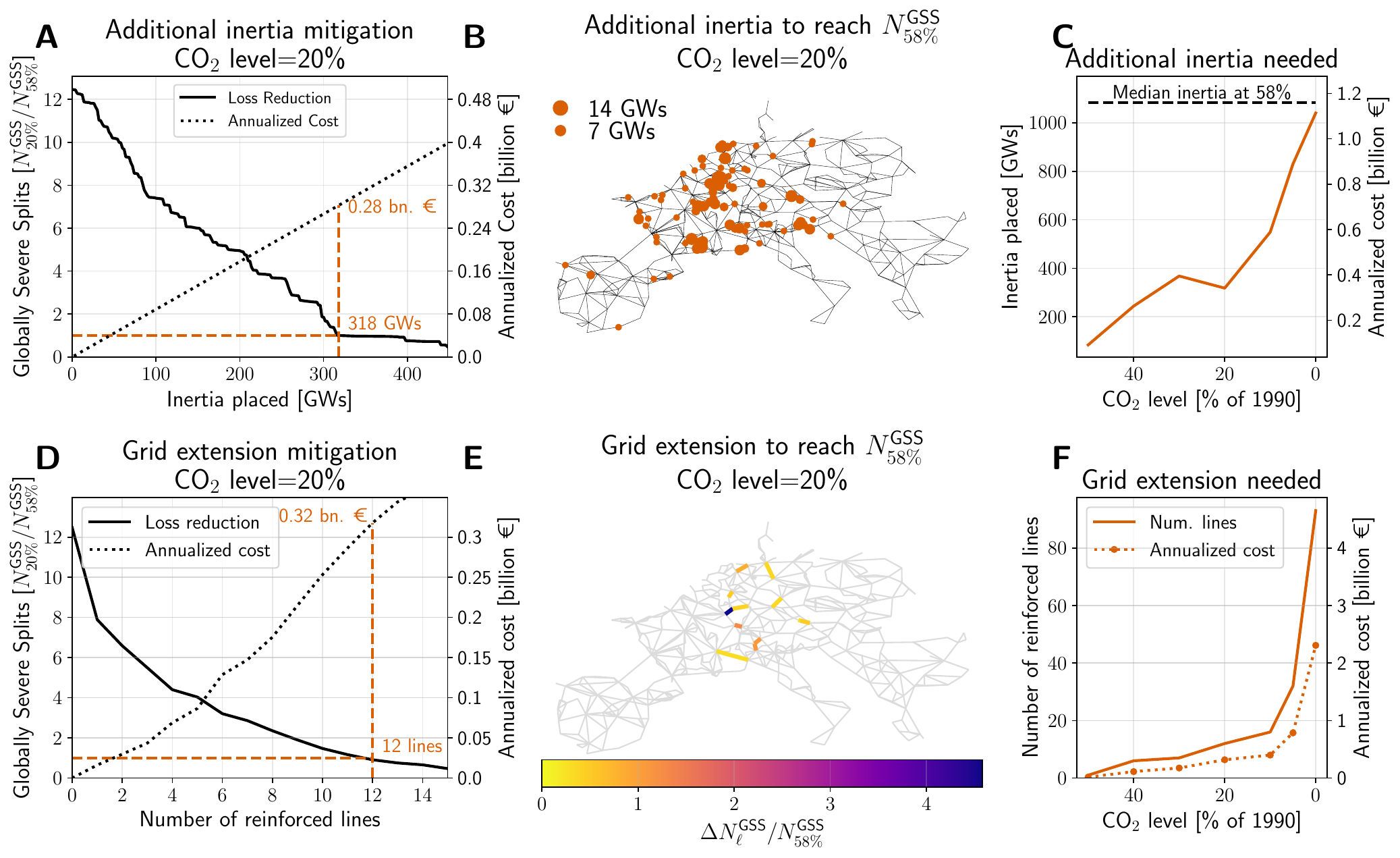}
\caption{
    \textbf{Mitigation of split-induced blackouts via additional inertia (top) and grid reinforcement reserved for stability margins (bottom).} Mitigation measures are applied independently using greedy algorithms (see Appendix).
    \textbf{a,} Effect of additional inertia at 20\% CO$_2$. The expected number of Global Severe Splits (GSS), conditioned on an $N-2$ contingency, is shown relative to the reference scenario ($R/R_{58\%}$) together with annualised costs. Achieving $R/R_{58\%}=1$ requires $318 \si{GWs}$ (orange lines).
    \textbf{b,} Optimal placement of 3$318 \si{GWs}$. Inertia concentrates in central Europe, with major contributions from France, northern Italy, Germany, Spain and the North Sea region.
    \textbf{c,} Additional inertia required to reach $R/R_{58\%}=1$ as a function of CO$_2$ level. Requirements rise sharply below 20\% emissions and approach the median inertia of the reference system at 0\%. The annualized cost which is directly proportional to the amount of inertia is shown on the secondary axis.
    \textbf{d,} Effect of grid reinforcement at 20\% CO$_2$, analogous to panel \textbf{a}. Reinforcing 12 lines restores $R/R_{58\%}=1$ at annualised costs of 0.14\,billion Euro (orange lines).
    \textbf{e,} Location of selected reinforcements, corresponding to major bottlenecks identified in Fig.~\ref{fig:line_failures}, including the France–Germany interface, the Alpine corridor and the North Sea region.
    \textbf{f,} Additional transmission lines required to reach $R/R_{58\%}=1$ and associated costs as a function of CO$_2$ level. Below 10\% emissions, required reinforcements increase markedly. Costs scale with total line length and thus approximately with the line number.
}
\label{fig:mitigation}
\end{figure*}

Rather than mitigating RoCoF violations after a split, blackout risk can also be reduced upstream by lowering the probability of cascading failures. Operating the existing grid more conservatively, by reserving additional transmission capacity beyond the $N-1$ criterion, would improve security but at the expense of reduced power exchanges and higher system costs. We therefore consider targeted reinforcement of selected transmission corridors, with all added capacity reserved explicitly for security rather than additional trading. The economic dispatch remains unchanged. This roughly corresponds to implementing an $N-2$ security constraint for selected vulnerable corridors, which represents a shift in operational and market design. 

Using a greedy selection algorithm to minimise the expected number of GSS (see Appendix), we identify cost-effective reinforcements (Fig.~\ref{fig:mitigation}). At 20\% CO$_2$, twelve additional $380 \, \si{kV}$ lines suffice to restore GSS risk to the reference level. These reinforcements concentrate in central transmission corridors, notably along the east–west interface at the German–French border, in the Alpine region, and in northern and southern Germany where major wind transfers occur. Even in the fully decarbonised scenario, annualised reinforcement costs remain below $2.5$ billion Euro. Relative to total system expenditures, these investments are modest, indicating that targeted grid reinforcement provides a cost-effective means of enhancing structural robustness.

\section{Discussion}

We quantify the risk of cascading transmission failures and system splits under cost-optimal European decarbonization pathways. Deep decarbonization does not merely alter the generation mix, but changes the structural conditions under which large-scale failures unfold. High shares of wind and solar reduce synchronous inertia while increasing long-distance transfers due to spatial mismatches between generation and load. The combination of lower inertia and higher structural flows increases the susceptibility to severe system splits if infrastructure and control strategies remain unchanged. Clustering analysis reveals persistent and emerging split patterns aligned with structurally critical transmission bottlenecks such as Alpine region, the Balkans and the Iberian Peninsula (Fig.~\ref{fig:cluster}), whose frequency and severity increase markedly at very low CO$_2$ levels.

Recent ENTSO-E assessments identify system splits as a major resilience risk in the European power system~\cite{entso-e_project_2025}. Their analyses quantify the survivability of predefined partitions at country-level resolution. Our analysis extends this perspective by explicitly modelling the formation of system splits through cascading transmission failures, starting from $N-1$ secure operating states. We find that most severe events involve multi-component fragmentation rather than simple two-island separations, and that split locations align with existing bottlenecks. Blackout risk is therefore structured by network topology and renewable deployment patterns rather than random contingencies.

Several mitigation pathways emerge from our analysis. Additional inertia directly addresses post-split frequency instability. In the fully decarbonized scenario, roughly $1000 \,\si{GWs}$ of additional inertia—comparable to the median inertia of today’s system—reduces Global Severe Split (GSS) risk to reference levels at annualized costs below 1.2 billion Euro. Beyond split prevention, additional inertia from synchronous condensers or grid-forming inverters can improve frequency quality in non-split disturbances and support voltage regulation, providing additional system-wide benefits.

Alternatively, blackout risk can be reduced upstream by limiting cascade formation. Transmission reinforcement, however, does not automatically improve robustness. If additional capacity is fully utilised for economic dispatch, it can increase long-distance transfers~\cite{50hertz2025sysstab} and thereby amplify structural imbalances. In our analysis, reinforcements reduce GSS risk because added capacity is reserved to enhance stability margins rather than to facilitate additional trading. Under this condition, targeted expansion of key corridors restores GSS risk to present-day levels at annualised costs below 2.5 billion Euro. Current regulatory frameworks generally prioritise full utilisation of available transmission capacity; reserving margins explicitly for robustness would therefore represent a shift in operational and market design. As inertia provision and grid expansion exhibit diminishing returns, a hybrid strategy combining selective reinforcement with targeted inertia deployment can be most cost-effective.

A third lever lies in system architecture itself. More balanced spatial deployment of wind and solar generation can reduce long-distance transfers and structural vulnerability~\cite{frysztacki2021-ae}. Implementing this approach, however, would require incentives or planning interventions that internalize grid robustness alongside cost minimization.

Overall, our findings demonstrate that decarbonization cannot be reduced to a question of generation technology alone. While public debate often centres on expanding renewable capacity, the spatial configuration of that expansion and the structure of the transmission network fundamentally determine system robustness. Transmission is not merely an enabling infrastructure but a structural component of system security, shaping how disturbances evolve and whether contingencies escalate into large-scale blackouts. Generation, transmission and control must therefore be co-designed rather than treated as independent dimensions of the transition. Designing a deeply decarbonized European power system that is both low-carbon and resilient requires an integrating transmission planning. Stability constraints and robustness considerations should be embedded directly into long-term system optimisation and operation.

\appendix

\section{Power system model and data}

Our study is based on PyPSA-Eur, an open-source model of the European power system~\cite{pypsa-website,horsch2018pypsa}. Hourly electricity demand and power plant data are obtained from the Open Power System Data project~\cite{OPSD-demand,opsd-conventionalpowerplants}. Transmission topology is derived from the ENTSO-E interactive map~\cite{interactive} using the GridKit toolkit~\cite{wiegmans_2016_55853}. Wind and solar generation time series follow Ref.~\cite{REatlas}. Cost assumptions for 2030 are based on projections from Refs.~\cite{schroeder2013,etip,dea2019,budischak2013}. We use weather data from 2013, which is considered representative~\cite{gotske2024designing}. The model includes the British Isles and Scandinavia but excludes Turkey; cascade simulations focus on the Central European synchronous area.

For computational tractability, simulations use three-hour time steps and an aggregated transmission network with 600 nodes and 1042 connection corridors. Nodes are clustered hierarchically by merging neighbouring regions with similar wind and solar capacity factors, preserving renewable variability and network topology~\cite{frysztacki2021-ae}. Generation, storage and transmission assets are aggregated accordingly.
All simulations employ the linearised (DC) power flow approximation and a DC optimal power flow framework~\cite{Wood14}.

\section{Scenarios and Power System Optimization}

PyPSA enables joint optimisation of investments and operation over one year to minimise total annualised system costs. Explicit inclusion of all $N-1$ security constraints in this optimisation is computationally intractable; we therefore adopt a two-stage approach.

In the first stage, investments and dispatch are optimised under a CO$_2$ emissions cap using a brownfield approach. $N-1$ security is approximated by requiring that no transmission line exceeds 70\% of its rated capacity. Load shedding is allowed but penalised with prohibitively high costs.

In the second stage, storage operation is fixed to the stage-one solution and generator dispatch is determined using a Security-Constrained Linear Optimal Power Flow (SCLOPF) that enforces explicit $N-1$ constraints at each time step. This replaces the 70\% loading heuristic with post-contingency flow constraints.

The simulations do not include an explicit unit commitment model, as this would require solving a large mixed-integer linear program. Hence, we approximate that a plant is online and provides inertia when its output exceeds 5\% of rated capacity. Many real plants have higher minimum operating points, this approximation can lead to a slight overestimation of the number of online power plants and the corresponding online inertia.


To ensure feasibility, two minor relaxations are introduced. First, generation and storage capacities are increased by 1\% relative to stage one. Second, a highly penalised artificial load is added at each node to resolve infeasibilities arising from fixed storage state-of-charge constraints.

\section{Simulation and evaluation of cascading failures}
\label{sec:methods:simulation}

For all scenarios and time steps, the grid is $N-1$ secure by construction. We therefore simulate all possible double line failures, excluding bridges (i.e. links whose removal trivially disconnects the network). In the aggregated network, nodes may be connected by multiple circuits, potentially at different voltage levels. Unique failure combinations are simulated and duplicates accounted for via a weighting factor. Power flows are evaluated using the DC approximation~\cite{Wood14}. Common-mode failures are not treated separately; all double contingencies are considered equally.

\begin{figure}[tb]
\caption{
\textbf{Algorithm for cascading failures for a single snapshot $T$ of the system.}\label{alg:split}}
\hrule
\small
\begin{algorithmic}
        \State Determine the power injections $\vec{\mathbf{P}}$
        \For{$\ell_i$ and $\ell_j$ in the set of all non-bridge lines}
            \State Trigger cascade by removing one circuit
            \State from each corridor $\ell_i$ and $\ell_j$
            \While{System is connected}
            \State Recalculate power flows $\vec{\mathbf{f}}$
            \If{any $| f_i | > f^{\mathrm{max}}_i$} 
                \State remove all overloaded lines
            \Else
                \State \textbf{break}
            
            \EndIf
            
            \If{System splits}
                \State Store component-level variables:
                \State -- load
                \State -- power imbalance $\Delta P(c)$ 
                \State -- rated power of inertia-providing plants $S_{on}(s)$
                \State \textbf{break}
            \EndIf
            \EndWhile
        \EndFor
\end{algorithmic}
\medskip
\hrule
\end{figure}

Cascading failures are simulated in a quasi-steady-state framework. At each step, real power flows are computed and overloaded lines are removed to represent protection actions. If there are additional circuits that connect the same aggregated nodes as an overloaded line, all are removed simultaneously. In the aggregated representation, such circuits are electrically identical and highly correlated in loading; sequential removal would introduce arbitrary ordering effects without materially altering the cascade outcome. The procedure continues until no further overloads occur or the grid separates. Note that this may result in multiple disconnected components forming within a single cascade step. Generation and HVDC injections are held constant during the cascade (cf.~Fig.~\ref{alg:split}).

If a split occurs, we record the power imbalance $\Delta P(c)$ and the rated capacity of online inertia-providing plants $S_{\mathrm{on}}(c)$ for each component $c$, and compute the RoCoF using Eq.~\eqref{eq:rocof}. The impact of a split is quantified by the fraction of load not served. For a split event $k$ at time $t$, a component $c$ is assumed to experience a complete blackout if $|\text{RoCoF}(c)| > 1\,\si{Hz/s}$, resulting in loss of all load within that component. The fraction of load not served is defined as
\begin{equation}
    r_k = \frac{\sum'_{i} L_{i,t}}{\sum_j L_{j,t}},
\end{equation}
where the primed sum runs over nodes $i$ in components with $|\text{RoCoF}(c)| > 1\,\si{Hz/s}$ and $L_{i,t}$ denotes the load at node $i$ and time $t$.

\begin{figure}
    \centering
    \includegraphics[width=1.0\linewidth]{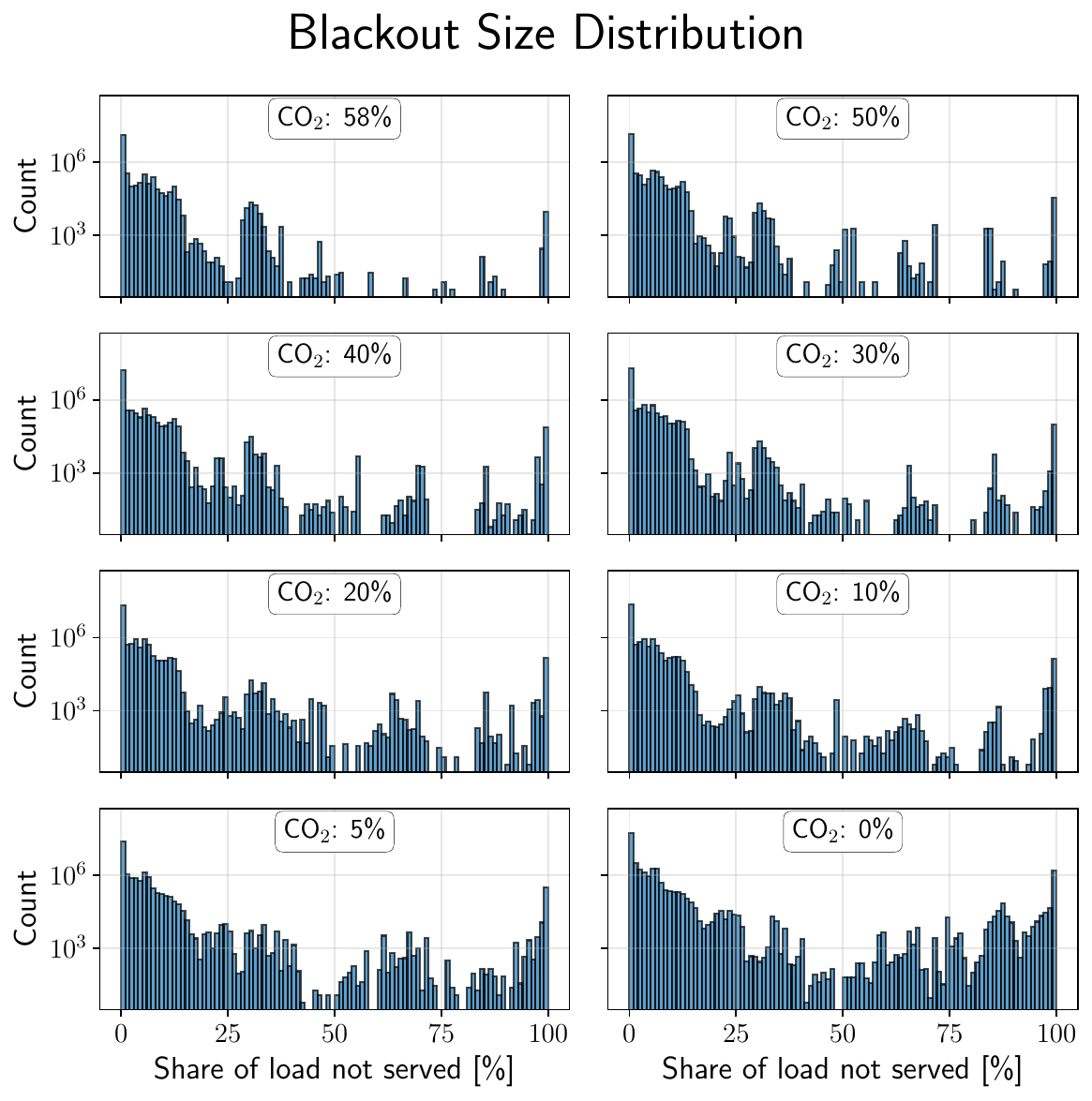}
    \caption{Blackout Size histograms for the different scenarios. The likelihood of all sizes increases during decarbonization. Medium sized blackouts (20\%-80\%) are generally unlikely.}
    \label{fig:blackout-sizes}
\end{figure}

For all scenarios system splits without any RoCoF violation are the most frequent, followed by small blackouts (Fig.~\ref{fig:blackout-sizes}). Large but not globally severe splits are generally unlikely since the large power imbalances that lead to a RoCoF violations in the majority of the grid are most likely accompanied by RoCof violations in all of the grid. For a small component to remain stable, it would have to have a higher relative inertia compared to its power imbalance than the rest of the grid. This is unlikely since it will in most cases have much higher power imbalance per node than the rest of the grid. Hence, a split with a blackout in one large component typically corresponds to a global severe split.

\section{Clustering of system split events}

To identify recurring blackout patterns, we cluster all system split events using a similarity metric that accounts for both the node-wise outage states and the geometry of the resulting system-split boundaries.
Each split event $k$ is represented by a node label vector $\mathbf{l}^{(k)} \in \{ -1, 0, +1 \}^N$, indicating blackout due to positive RoCoF violation, stable operation, or blackout due to positive RoCoF violation, respectively. 
Similarly, we define a binary edge-outage vector $\mathbf{e}^{(k)} \in \{0,1\}^{|E|}$ for visualization purposes, indicating which transmission lines fail during event $k$.
We concentrate our analysis on large scale blackouts and discard events resulting in less than $10\%$ load not served. After removing duplicates we are left with 66458 unique node label vectors.

In order to cluster events, we define a metric for their (dis-)similarity by adapting the hamming distance. For two events $k$ and $n$, node-wise agreement is quantified by the normalized categorical similarity
\begin{equation}
s_{\mathrm{ham}}(k,n)
= 1 - \frac{1}{N} \sum_{i=1}^N \mathbf{1}\left[ l^{(k)}_i \neq l^{(n)}_i \right],
\end{equation}
where $\mathbf{1}$ denotes the indicator function.

System splits create interfaces where neighboring nodes have different labels.  
Using the incidence matrix $B$, we identify cut edges through
\begin{equation}
\mathbf{c}^{(k)} = \mathbf{1}\!\left[ B^{\mathsf T}\mathbf{l}^{(k)} \neq 0 \right] \in \{0,1\}^{|E|},
\end{equation}
and nodes adjacent to at least one cut edge form the binary boundary mask
\begin{equation}
\mathbf{b}^{(k)}_{(0)} = \mathbf{1}\!\left[ \left| \mathbf{c}^{(k)\mathsf T} B \right| > 0 \right] \in \{0,1\}^N .
\end{equation}

Small geometric shifts in the cut set should not dominate the similarity measure, so we smooth the boundary mask using a short diffusion on the network.
 
With the random-walk matrix $P = D^{-1}A$, where $A$ is the adjacency and $D$ the degree matrix, we iterate
\begin{equation}
\mathbf{b}^{(k)}_{(t+1)}
= (1-\alpha)\,\mathbf{b}^{(k)}_{(t)} + \alpha\, P\,\mathbf{b}^{(k)}_{(t)},
\qquad t=0,\dots,T-1 ,
\end{equation}
with $\alpha=0.4$ and $T=3$. In the resulting continuous field
$\mathbf{b}^{(k)} = \mathbf{b}^{(k)}_{(T)} \in [0,1]^N$, 
large entries correspond to nodes close to the system-split boundary.

We compute the geometric similarity of two split boundaries using the cosine similarity of their smoothed fields,
\begin{equation}
s_{\cos}(k,n) =
\frac{ \langle \mathbf{b}^{(k)}, \mathbf{b}^{(n)} \rangle }
     { \|\mathbf{b}^{(k)}\|_2 \, \|\mathbf{b}^{(n)}\|_2 }
.
\end{equation}

Finally, we use the geometric mean to combine node-wise agreement and geometric boundary similarity into a single similarity score
\begin{equation}
s(k,n)
= \sqrt{s_{\mathrm{ham}}(k,n)\; s_{\cos}(k,n)} \in [0,1]
\end{equation}
This metric equals one only for identical outage patterns and captures both node level outcomes and where the system splits occur.  
It provides a robust basis for clustering blackout events with similar spatial failure structures. The combined metric is necessary because nodal agreement alone can be misleading.  

Two split events may assign the same class to the vast majority of nodes while still differing substantially in where the system actually separates.  
Clusters~7 and~9 in Fig.~\ref{fig:cluster} illustrate this: most nodes share the same class, yet the spatial structure of the split is entirely different.  
The geometric boundary similarity prevents such cases from being grouped together by ensuring that events are considered similar only when both their node labels and the location of the system split agree.

We tested multiple agglomerative clustering configurations and chose 96 clusters, as this setting produced the highest silhouette score.
Each cluster $C$ is characterized by a centroid defined as the weighted mean of its three-class node-label vectors, yielding a $3 \times N$ probability distribution over node classes.
Similarly, we compute the mean of the binary edge outage vectors assigned to a cluster to represent typical cascade paths. We rank clusters by their relative frequency $\beta$, and their contribution to the overall share of load not served, which is defined as
\begin{equation}
    R(C) = \frac{\sum_{k \in C} r_k}{\sum_k r_k} .
\end{equation}

\begin{figure}[tb]
\caption{\textbf{Greedy synthetic inertia placement.} \\
$comps$: set of split components with their properties. \\
$\Delta m$: inertia increment [GW]. \\
$nodes$: set of candidate nodes for inertia placement. \\
$i_{\max}$: maximum number of placement steps. \\
$n_{GSS}$: list of total number of GSS at each step. \\
$n_{\rm placement}$: list of selected nodes for each step. \\
$\texttt{N\textsubscript{GSS}}(comps)$: computes the total number of GSS from component data. \\
$\texttt{ADD\_INERTIA}(comps, \Delta m, n)$: returns updated $comps$ with $\Delta m$ added at node $n$. \\
$\texttt{ARGMIN}(x)$: index of the smallest element of array $x$.
}
\label{alg:inertia}
\hrule
\small
\begin{algorithmic}
\Procedure{placement}{$comps$, $\Delta m$, $nodes$, $i_{\max}$}
    \State $n_{GSS} \gets [\texttt{N\textsubscript{GSS}}(comps)]$
    \State $n_{\rm placement} \gets [\varnothing]$
    \State $i \gets 1$
    \While{$i \leq i_{\max}$}
        \State $n_{GSS,\rm post} \gets$ array of length $\text{len}(nodes)$
        \For{$n$ in $nodes$}
            \State $comps_n \gets \texttt{ADD\_INERTIA}(comps, \Delta m, n)$
            \State $n_{GSS,\rm post}[n] \gets \texttt{N\textsubscript{GSS}}(comps_n)$
        \EndFor
        \State $n_{GSS, \rm opt} \gets \min(n_{GSS,\rm post})$
        \State $n_{\rm opt} \gets \texttt{ARGMIN}(n_{GSS,\rm post})$
        \If{$n_{GSS,\rm opt} < n_{GSS}[i-1]$}
            \State $comps \gets \texttt{ADD\_INERTIA}(comps, \Delta m, n_{\rm opt})$
            \State append $n_{GSS,\rm opt}$ to $n_{GSS}$
            \State append $n_{\rm opt}$ to $n_{\rm placement}$
            \State $\Delta m \gets 1$
            \State $i \gets i+1$
        \Else
            \State $\Delta m \gets 2 \cdot \Delta m$
        \EndIf
    \EndWhile
    \State \Return $n_{GSS}$, $n_{\rm placement}$
\EndProcedure
\end{algorithmic}
\hrule
\end{figure}

\begin{figure}[tb]
\caption{\textbf{Heuristic for transmission line reinforcement.}\\
\textit{events}: set of cascade events with associated lost load.\\
\textit{lines}: set of candidate transmission lines.\\
$\texttt{cost}(l)$: investment cost of reinforcing line $l$ (proportional to length).\\
$\texttt{is\_GSS}(e)$: returns true if event $e$ leads to a GSS. \\
$n_{GSS}$: list of total number of GSS at each step. \\
$\texttt{N\_GSS}(l)$: number of GSS triggered by line $l$.\\
$\texttt{triggers}(e)$: pair of trigger lines of event $e$.\\
$n_{\mathrm{GSS}}^{\mathrm{ref}}$: reference number of GSS (58\% scenario).\\
$\textit{ratio}(l)$: ratio of cost of reinforcement to number of mitigated GSS for line $l$.\\
$\texttt{reinforced\_lines}$: list of reinforced lines.
}
\label{alg:reinforce}
\hrule
\begin{algorithmic}[1]
\Procedure{Reinforce}{$events, lines, n_{\mathrm{GSS}}^{\mathrm{ref}}$}
    \State $n_{\mathrm{GSS}} \gets [\sum_{e \in events} \texttt{is\_GSS}(e)]$
    \State $\texttt{reinforced\_lines} \gets [\varnothing]$
    \While{$n_{\mathrm{GSS}} > n_{\mathrm{GSS}}^{\mathrm{ref}}$}
        \ForAll{$l \in \textit{lines}$}
            \State $\texttt{N\_GSS}(l) \gets \sum_{e:\, l \in \texttt{triggers}(e)} \texttt{is\_GSS}(e)$
            \If{$\texttt{N\_GSS}(l)=0$}
                \State $ratio(l) \gets +\infty$
            \Else
                \State $ratio(l) \gets \texttt{cost}(l) / \texttt{N\_GSS}(l)$
            \EndIf
        \EndFor
        \State $l_{\mathrm{opt}} \gets \arg\min_{l \in \textit{lines}}\, ratio(l)$
        \State append $l_{opt}$ to $\texttt{reinforced\_lines}$
        \State remove all $e$ with $l_{\mathrm{opt}} \in \texttt{triggers}(e)$ from $events$
        \State append $\sum_{e \in events} \texttt{is\_GSS}(e)$ to $n_{\mathrm{GSS}}$
    \EndWhile
    \State \Return $n_{\mathrm{GSS}}$, $R$
\EndProcedure
\end{algorithmic}
\hrule
\end{figure}

\section{Additional inertia}

Additional inertia, e.g.~from synchronous condensers or virtual inertia, can reduce the RoCoF following a system split and thereby prevent blackouts. As our objective is to prevent Global Severe Splits (GSS), we restrict the optimisation to these events. Placement of additional inertia is determined using a greedy search algorithm. In each iteration, a fixed increment $\Delta m$ is added to the node $n_{\rm opt}$ that maximises the reduction in aggregated load not served. Larger increments are used at lower emission levels to limit computational effort. If multiple nodes yield identical reductions, one of them is selected at random. If no node reduces lost load, $\Delta m$ is doubled until an improvement is achieved. Further details are provided in Fig.~\ref{alg:inertia}. Unlike approaches based on predefined split scenarios or individual case studies~\cite{50hertz2025sysstab,knechtges2023identification,entso-e_project_2025}, our optimisation accounts for all simulated GSS events.

\section{Grid extensions and increased stability margins}

Increasing security margins beyond the $N-1$ criterion can reduce the risk of severe cascading failures, but at the cost of reduced transmission capacity and higher system costs. We therefore analyse a scenario in which selected transmission corridors are reinforced to increase redundancy during cascades. The added capacity is assumed to be reserved for security margins rather than additional trading, such that dispatch remains unchanged.

Reinforcements are identified using a heuristic approach. Adding one circuit to an existing corridor implies that a double failure involving the reinforced line is equivalent to a single-line outage in the original grid. Since the original system is $N-1$ secure, such contingencies cannot trigger cascades. We therefore approximate that reinforcing a corridor eliminates cascades initiated by its failure. 
We note that this reinforcement can also decrease the severity of other cascades by preventing secondary failures. Since our approach can not capture this effect, our results represent a lower bound regarding the effectiveness of grid reinforcements.

Corridors are selected using an iterative greedy algorithm restricted to GSS events. At each step, we evaluate the reduction in load not served achieved by reinforcing each corridor and select the one maximising the ratio of prevented load loss to investment cost. Further details are provided in Fig.~\ref{alg:reinforce}. Investment costs are proportional to line length with $4.5 \times 10^6$ \si{EUR/km} based on estimates from the Netzentwicklungsplan~\cite{noauthor_kostenschatzungen_nodate}.


\end{document}